\documentclass[10pt]{IEEEtran}
\usepackage[utf8]{inputenc}
\usepackage[T1]{fontenc} 

\usepackage{tikz}
\usetikzlibrary{svg.path}
\definecolor{orcidgreen}{HTML}{A6CE39}
\newcommand\orcid[1]{\href{https://orcid.org/#1}{\raisebox{-.15em}{\resizebox{1em}{1em}{%
      \begin{pgfpicture}
        \pgfsetcolor{orcidgreen}
        \pgfpathsvg{M106.301,35.435c0,19.572-15.863,35.436-35.436,35.436S35.43,55.007,35.43,35.435S51.293,0,70.865,0
	S106.301,15.863,106.301,35.435z}
        \pgfusepath{fill}
        \pgfsetcolor{white}
        \pgfpathsvg{M59.321,32.39v13.398h-4.264V16.14h4.264V32.39z}
        \pgfpathsvg{M65.578,16.112h11.572c9.772,0,15.724,7.226,15.724,14.839c0,7.004-4.816,14.838-15.779,14.838H65.578 V16.112z M69.841,41.941h6.562c8.305,0,12.098-5.039,12.098-10.99c0-3.654-2.215-10.99-11.877-10.99h-6.782V41.941z}
        \pgfpathsvg{M57.189,54.758c-1.55,0-2.796-1.245-2.796-2.796c0-1.523,1.246-2.797,2.796-2.797s2.796,1.273,2.796,2.797 C59.985,53.485,58.74,54.758,57.189,54.758z}
        \pgfusepath{fill}
      \end{pgfpicture}%
  }}}}

\usepackage[normalem]{ulem}

\usepackage{amsmath,amssymb,amsfonts}
\usepackage{pgffor}
\usepackage{xcolor}
\usepackage{xifthen}
\usepackage{xargs}

\usepackage{newfloat}
\usepackage{syntax}
\usepackage{xpatch}

\usepackage{caption}    
\usepackage{listings}
\usepackage{xspace}
\usepackage{graphicx}
\usepackage{calc}
\usepackage{ragged2e}
\usepackage{scrextend}
\usepackage{changepage}
\usepackage{bm}
\usepackage{comment}
\usepackage{booktabs}
\usepackage{array}

\usepackage{hyperref}

\definecolor{codeblue}{rgb}{0.2,0.2,1}
\definecolor{codegreen}{rgb}{0,0.6,0}
\definecolor{codegray}{rgb}{0.5,0.5,0.5}
\definecolor{codepurple}{rgb}{0.58,0,0.82}
\definecolor{codeorange}{rgb}{.8,0.3,0.1}
\definecolor{codebggray}{rgb}{0.935,0.940,0.945}

\makeatletter
\lst@InstallKeywords k{assert}{assertstyle}{}{basicstyle}{}ld
\makeatother

\makeatletter
\lst@InstallKeywords k{adt}{adtstyle}{}{basicstyle}{}ld
\makeatother

\makeatletter
\def\lst@Literatekey#1\@nil@{\let\lst@ifxliterate\lst@if
    \expandafter\def\expandafter\lst@literate\expandafter{\lst@literate#1}}
\makeatother

\newcommand{\makechar}[1]{
  \begingroup
    \fontencoding{T1}%
    \fontfamily{qcr}% TeX Gyre Cursor
    \selectfont
    \unskip % removes space before char
    \string#1%
  \endgroup
}

\lstdefinestyle{pystyle}{
    language=Python,
    literate={\ \ \ \ }{{\ \ }}{2},
    literate={^}{\makechar{^}}{1},
    basicstyle=\small\ttfamily,
    commentstyle=\color{codegreen},
    keywordstyle=\color{codeblue}\bfseries,
    numberstyle=\tiny\color{codegray},
    stringstyle=\color{codepurple},
    adtstyle=\color{codeorange}\bfseries,
    moreadt={Sum,Product,Enum,TaggedUnion},
    tabsize=2,
    deletekeywords=[2]{format,sum,abs,float},
    morekeywords={assert,with,as},
    showstringspaces=false,
}

\lstdefinestyle{inline}{
    style=pystyle,
    basicstyle=\ttfamily,
    deletekeywords={and},
    backgroundcolor={codebggray},
}

\DeclareCaptionLabelFormat{center}{\hspace*{\fill}#1 #2\hspace*{\fill}}

\DeclareCaptionStyle{capstyle}{
    style=base,
    textfont={normal,small},
    labelfont=bf,
    margin=.01\columnwidth,
}

\DeclareFloatingEnvironment[name=Example]{examplec}
\DeclareFloatingEnvironment[name=Algorithm]{algorithm}
\counterwithin{examplec}{section}
\newcommand{\VARIADIC}[6]{%
    \expandafter\newcommand\csname Check#1Arg\endcsname[1]{%
        \csname @ifnextchar\endcsname\bgroup{%
            \csname r#1Case\endcsname{##1}%
        }{%
           #2{##1}#6%
        }%
    }%
    \expandafter\newcommand\csname i#1Case\endcsname[1]{%
        \csname Check#1Arg\endcsname{#4{##1}}%
    }%
    \expandafter\newcommand\csname r#1Case\endcsname[2]{%
        \csname Check#1Arg\endcsname{#5{##1}{##2}}%
    }%
    \expandafter\newcommand\csname #1\endcsname[0]{%
        \csname @ifnextchar\endcsname\bgroup{%
            \csname i#1Case\endcsname%
        }{%
            #2{#3}#6%
        }%
    }%
}

\makeatletter
\providecommand\usr@lambda{}
\makeatother
\newcommand{\lambdac}[2][0]{%
    \expandafter\renewcommand\csname usr@lambda\endcsname[#1]{#2}%
    \csname usr@lambda\endcsname%
}
\VARIADIC{tocite}{(\textcolor{red}}{cite}{cite:~}{\lambdac[2]{#1,~#2}}{)}
\newcommand{\peak}[0]{PEak\xspace}
\newcommand{\magma}{Magma\xspace}

\newcommand*{\code}{\lstinline[mathescape=true,style=inline]}

\newcommand*{\dunder}[1]{\code{__#1__}}

\newcommand*{\bvs}[0]{BitVector}

\newcommand*{\bvc}[1][]{%
    \ifthenelse{\isempty{#1}}%
        {\text{\bvs}}%
        {\text{\bvs[#1]}}%
}
\newcommand*{\bv}[1][]{%
    \code?$\bvc[#1]$?\xspace%
}
\newcommand{\bit}[0]{\code{Bit}\xspace}
\newcommand{\hwt}[0]{\texttt{hwtypes}\xspace}
\newcommand{\astt}[0]{\texttt{ast\_tools}\xspace}

\newcommand{\sigbrief}[1]{%
    \foreach \T [count=\i] in {#1} {%
        \ifnum\i=1%
            \text{\T}%
        \else%
            \rightarrow \text{\T}%
        \fi%
    }%
}

\newcommand{\vpre}[0]{}%\vspace{-0.5em}}
\newcommand{\vpost}[0]{\vspace{-0.25em}}
\newcommand{\vhead}[0]{}%\vspace{-0.05em}}
\newcommand{\vpresub}[0]{}%\vspace{-0.05em}}

\title{\peak: A Single Source of Truth for Hardware Design and Verification}

\makeatletter
\newcommand{\linebreakand}{%
  \end{@IEEEauthorhalign}
  \hfill\mbox{}\par
  \mbox{}\hfill\begin{@IEEEauthorhalign}
}
\makeatother
\author{%
    \IEEEauthorblockN{
        Caleb Donovick\IEEEauthorrefmark{1}\orcid{0000-0001-9336-1267},
        Ross Daly\IEEEauthorrefmark{1}\orcid{0000-0002-4938-5250},
        Jackson Melchert\IEEEauthorrefmark{2}\orcid{0000-0002-8232-1603},
        Lenny Truong\IEEEauthorrefmark{1}\orcid{0000-0001-7583-9730},\\
        Priyanka Raina\IEEEauthorrefmark{2}\orcid{0000-0002-8834-8663},
        Pat Hanrahan\IEEEauthorrefmark{1}\orcid{0000-0002-3474-9752}, %
        and Clark Barrett\IEEEauthorrefmark{1}\orcid{0000-0002-9522-3084}
    } \\
    \IEEEauthorblockN{$ $ } \\
    \IEEEauthorblockA{%
        \IEEEauthorrefmark{1}Dept. of Computer Science:
        {\{donovick, rdaly525, lenny, hanrahan, barrett\}@cs.stanford.edu}\\
        \IEEEauthorrefmark{2} Dept. of Electrical Engineering:
        {\{melchert, praina\}@stanford.edu} \\
        Stanford University, Stanford CA, USA
    }
}

\LetLtxMacro{\oldsection}{\section}
\LetLtxMacro{\oldsubsection}{\subsection}
\LetLtxMacro{\oldsubsubsection}{\subsubsection}
\renewcommand{\section}[2][]{\oldsection{#2}\label{sec:#1}\vhead}
\renewcommand{\subsection}[2][]{\vpresub\oldsubsection{#2}\label{sec:#1}\vhead}

\begin{document}

\maketitle

\begin{abstract}
Domain-specific languages for hardware can significantly en\-hance designer productivity, but sometimes at the cost of ease of verification. On the other hand, ISA specification languages are too static to be used during early stage design space exploration. We present \peak, an open-source hardware design and specification language, which aims to improve both design productivity and verification capability. \peak does this by providing a single source of truth for functional models, formal specifications, and RTL. \peak has been used in several academic projects, and \peak-generated RTL has been included in three fabricated hardware accelerators.  In these projects, the formal capabilities of \peak were crucial for enabling both novel design space exploration techniques and automated compiler synthesis.
\end{abstract}

\section[intro]{Introduction}

Domain-specific languages (DSLs) for hardware allow designers to build generators that are impossible to express using traditional hardware description languages such as SystemVerilog and VHDL~\cite{bachrach2012chisel,truong2019golden}. Such generators are of increasing importance as specialized chips become the norm in a post-Dennard-scaling world~\cite{hennessy2019new,shacham2010rethinking}.  DSLs can also provide better correctness guarantees through type safety (a well-known pain point in Verilog). These factors have led to an explosion of new DSLs for hardware design over the last decade~\cite{bachrach2012chisel,durst2020type,koeplinger2018spatial,lockhart2014pymtl,truong2019golden}.

Unfortunately, the design of most hardware DSLs has not sufficiently taken into account the impact on verification~\cite{lockhart2018tpu}.  For example, using a Verilog simulator to debug DSL-generated designs is notoriously difficult, as information is lost or obscured during the compilation process.
A first step towards addressing this challenge is to include support for writing properties that can be translated to system verilog assertions (SVAs), and indeed several languages provide this (e.g., Chisel~\cite{dobis2023verification} and \magma~\cite{truong2020fault}).  More ambitious efforts aim to enable source-level debugging~\cite{zhang2022bringing}, which will likely be crucial for effective debugging of generated RTL, especially at later design stages.

On the other hand, DSL models are well-po\-si\-tioned to dramatically improve the \emph{early-stage} verification experience.  In particular, they can be leveraged to greatly improve debugging and verification during design exploration. Traditionally, separate functional models play a key role during this phase, but a promising alternative supported by some DSLs (e.g., pyMTL~\cite{lockhart2014pymtl}) is to automatically extract a high-performance executable functional model from a DSL description. Moreover, with the right semantics and support, the user can even be provided with direct access to an automatically-generated \emph{formal} model for the design, enabling novel and early uses of formal methods during the design exploration process.  Current DSLs provide very limited support for such features.

%One alternative approach for addressing verification challenges in the context of ISAs is to develop formal languages for ISA specification such as SAIL~\cite{sail2015}, ILA~\cite{ila2018}, and ISA-Formal~\cite{isaformal2016}. These are powerful tools, but they cannot be used to generate RTL.  While this disconnect makes sense when verifying new RTL against an existing ISA specification, it is tedious when the ISA itself being developed, as for each new candidate ISA, both its RTL and its specification must be written separately.
%CB The above paragraph feels out of place - maybe we can mention this if we talk about ISAs later?

In this paper, we introduce \peak, a Python-embedded DSL, with an accompanying set of open-source tools, including a compiler.  \peak provides a \emph{single source of truth} for compilation to RTL, functional simulation, and formal modeling.
Designers who use \peak do not need to implement the same thing multiple times, and the different implementations are guaranteed to be consistent with each other.  Furthermore, these capabilities directly enable novel formal-in-the-loop design methodologies.

\peak is partly motivated by work being done at the Stanford Agile Hardware center~\cite{aha2020dac,ahavlsi,ahahotchips,aha2023tecs},\footnote{\url{aha.stanford.edu}} where it has been used to generate coarse-grained reconfigurable array (CGRA) architectures\footnote{CGRAs~\cite{cgra3,cgra1,cgra2} are a spatial architecture similar to FPGAs and are composed of processing element (PE) and memory tiles, and a configurable routing network.} for three generations of chips, two of which were fabricated. Section~\ref{sec:eval} explains how the formal model generated by \peak was used to synthesize compiler components for different candidate architectures, thereby enabling a systematic and automatic exploration of the design space.

The rest of this paper is organized into the following sections: Section~\ref{sec:lib} describe \hwt and \astt which peak is built on; Section~\ref{sec:peak} describes the \peak language and how it can be extended; and Section~\ref{sec:eval} evaluates \peak as a tool for DSE, showing it can generate both high performance RTL as well as SMT models which are usable in a formal-in-the-loop design flow.  We discuss related work and conclude in Sections \ref{sec:related} and \ref{sec:conclusion}, respectively.

\section[lib]{Hardware Types and AST-Tools}
%\ready
%\cdnote{2.25 pages}

In this section, we introduce two libraries which serve as the foundation of \peak.  First, we discuss \hwt,\footnote{https://github.com/leonardt/hwtypes} which serves as both the type system and compilation target for \peak.  Second, we discuss \astt,\footnote{https://github.com/leonardt/ast_tools} an open-source package for Python abstract syntax tree (AST) analysis and rewriting, which is used both to build the \peak compiler and to extend \peak's meta-programming facilities. These libraries are independent of \peak, and may be of interest on their own.

\subsection[lib:hwtypes]{Hardware Types}
%\ready

The core of \peak is the python-embedded expression language of \hwt. \hwt provides a uniform interface for: functional simulation, via direct execution in Python; formal analysis, via automatic translation to formulas in the language of satisfiability modulo theories (SMT)~\cite{BSST21}; and RTL generation, via a compiler to \magma.  By unifying these types we ensure the equivalence of the generated functional, formal, and RTL models. 

\hwt defines abstract interfaces and type constructors for a number of types and kinds. This includes a \bit (Boolean) type, fixed-width \bv types (signed and unsigned), arbitrary-precision floating-point types, and algebraic data types (ADTs).  We first focus on the \bit and \bv types (we discuss the use of ADTs in Section~\ref{sec:peak:adt}). \bit types provide the usual Boolean operators: and \code`&`, or \code`|`, xor \code`^`, and not \code`~`; equality operators: equals \code`==`, and not equals \code`!=`;  and an \code{ite} (if-then-else) method.  

% \begin{table}[]
%     \centering
%     \begin{tabular}{c|c|c}
%          operator & python method name & SMT-Lib equivalent \\
%          \hline \hline 
%          \code?&?   & \dunder{and}    & $and$ \\
%          \code?|?   & \dunder{or}     & $or$ \\
%          \code?^?   & \dunder{xor}    & $xor$ \\
%          \code?~?   & \dunder{invert} & $not$ \\
%          \code?==?  & \dunder{eq}     & $=$*   \\
%          \code?!=?  & \dunder{ne}     & $distinct$* \\
%                     & \code?ite?      & $ite$ \\
%          \hline
%     \end{tabular}
%     \caption{Methods on bit types}
%     \small{* The $=$ and $distinct$ operators are parametric in SMT-LIB; the corresponding operators on \bit are defined only on \bit.}
%     \label{tab:bit}
%     \cdnote{this seems a bit verbose maybe I just do it in text}
% \end{table}

The SMT-LIB standard~\cite{BarFT-SMTLIB} defines a large set of arithmetic and bitwise functions on bitvectors.
%These functions are defined in both the base theory (\code{FixedSizeBitVectors}) and in its associated logics (\code{BV} and \code{QF_BV}). 
%CB: Unnecessary detail
The \hwt \bv interface defines a method for each of these functions.
%with the exception of \code{bv2nat}.  
%CB: actually bv2nat is not part of SMT-LIB, it's just a helper to explain the semantics.
For instance, the equivalent of the SMT-LIB term \code{(bvadd x y)} (bitvector addition), where \code{x} and \code{y} are of sort \code{(_ BitVec 16)},  or 16-bit bitvectors, is the hwtypes expression \code{x.bvadd(y)}, where \code{x} and \code{y} are of the type \bv[16]. More generally, if \code{f} is a function over bitvectors defined by SMT-LIB, then there is an equivalent method named \code{f} on the \hwt \bv type. As a convenience, these methods are also defined by overloading Python operators when appropriate.  For example: \code{x.bvadd(y)} can be invoked with \code{x + y}.  The semantics of sign-dependent operators are defined by their type. For example, \code{x < y} invokes \code{x.bvslt(y)} (signed less than) for signed \code{x} and \code{x.bvult(y)} (unsigned less than) for unsigned \code{x}. 

There are three implementations of the \bv and \bit types.  The first implementation is a pure Python functional model over constant values. The second wraps pySMT~\cite{pysmt2015} to generate SMT terms. Finally, Magma provides a third implementation which allows for the definition of circuits. This uniform interface allows for the same hwtypes program to be interpreted in multiple ways.  The pure Python implementation is used to simulate a circuit, the SMT implementation is used to generate a formal model, and the Magma implementation is used to generate actual RTL.

The real power of \hwt comes from its embedding in Python which facilitates the generation of complex formulas. For example, we can generate an adder tree over any number of inputs with the use of a recursive function as shown in Example \ref{code:hwt:0}. This can be easily generalized to perform reduction over any function as shown in Example \ref{code:hwt:1}.
\begin{examplec}
\begin{lstlisting}
def tadd(*args):
    n = len(args)
    if n == 0:
        return 0
    elif n == 1:
        return args[0]
    else:
        left = tadd(*args[:n//2])
        right = tadd(*args[n//2:])
        return left + right
\end{lstlisting}
\vpre
\caption{Adder tree generator.}
\label{code:hwt:0}
\begin{lstlisting}
def treduce(f, ident, *args):
    n = len(args)
    if n == 0:
        return ident
    elif n == 1:
        return args[0]
    else:
        largs = args[:n//2]
        rargs = args[n//2:]
        l = treduce(f,ident,*largs)
        r = treduce(f,ident,*rargs)
        return f(l, r)
\end{lstlisting}
\vpre
\caption{Reduction tree generator.}
\label{code:hwt:1}
\vpost
\end{examplec}

%Despite its name, \hwt can be used outside of hardware contexts.  \hwt provides a pythonic interface to pysmt which is useful for constraint solving generally.  In contrast to other python front-ends to SMT solvers, \hwt interoperates well with python's built in reflection methods, \code{isinstance} and \code{issubclass}.  For example, in pySMT, all terms, regardless of sort, have the type \code{FNode}; this means a developer must learn pySMT's specific reflection methods.  In \hwt, a term of the sort \code{(_ BitVec 16)} would have the the python type \bv[16].

It is important to note that \hwt is an expression language only; all statements are executed in pure Python following typical Python semantics.  This is in contrast to \peak (see Section~\ref{sec:peak}, below), which breaks away from the semantics of pure Python and reinterprets the meaning of if statements as \code{ite}s using AST rewriting. 

\subsection[lib:ast]{AST Tools}
%\ready
In order to be able to reinterpret Python code, we developed the \astt library, which provides a generic infrastructure for composing passes that analyze and transform the Python abstract syntax tree (AST).  The design is the result of our experience developing ad hoc AST rewrites for various DSLs (including \peak) and recognizing the need for a common infrastructure to serve these languages.

%% \peak's compiler is written using \astt, which is a general framework for python AST introspection and rewriting.  \astt allows a developer to decorate functions and classes with a series of passes.  The decorator handles the complications of capturing the environment, parsing the AST of the decorated code, and executing the rewritten code.  
 
\begin{examplec}
\begin{center}
\begin{lstlisting}
@apply_passes([loop_unroll()])
def foo():
    for i in unroll([1,3,9]):
        print(i)
\end{lstlisting}
\begin{lstlisting}
def foo():
    print(1)
    print(3)
    print(9)
\end{lstlisting}
\end{center}
\vpre
\caption{Code with loop unrolling applied.}
\label{code:ast}
\vpost
\end{examplec}

\subsubsection{Pass Architecture} The entry point to the \astt library is the \texttt{apply_passes} function, which takes a list of passes to run and returns a decorator that is used to transform a function or class.  The \texttt{apply_passes} function provides a generic prologue and epilogue, which handles logic common to most code transformers.  The prologue parses the marked code into an AST
%strips the \texttt{apply_passes} decorator so that the function is not invoked in an infinite loop,
and captures a closure of the environment.  The epilogue serializes the transformed AST into code and executes it using the captured environment.
Passes use a generic interface that consumes as arguments the current AST, the current environment, and a metadata dictionary.  A pass may modify any or all of these and return them as results to be used for the next pass or for the epilogue.

In addition to the pass infrastructure, \astt provides several useful utilities such as the ability to generate a \emph{free} name in the environment, which allows new variables to be introduced without clobbering existing mappings.  It also includes a collection of generic transformation and visitor passes that perform common operations.

% While this \emph{branch-less} version of the program would be possibly incorrect to execute for general software (side-effects may be incurred  from branches that are not taken), it provides a convenient form for constructing \peak representations.  In particular, for both the hardware description and the formal model, the logic in both branches of a conditional should be included.  %The result is then selected via a multiplexer or an \code{ite} term, respectively.

%Unlike general software, \peak does not need to consider side-effects.  With this in mind, \peak leverages the use of symbolic execution of the SSA program to construct an expression graph where \texttt{phi} nodes correspond to a selector (multiplexer).  Evaluating the SSA program with the formal types will produce a formal expression, while hardware types produce a hardware description. 
%CB: unclear that this is adding a lot - cut for space

\subsubsection{Macros} The macro sub-package provides a simple mechanism for performing syntactic rewrites of the Python AST.  When an explicit macro identifier is encountered, such as \texttt{unroll} in Example~\ref{code:ast}, the corresponding transformation is invoked (\texttt{loop_unroll}).  \peak employs the macro pattern to allow staged expansion of the specification.  For example, \texttt{if} statements marked as macros will be evaluated before they are compiled, allowing the user to distinguish between conditional logic intended to describe the generation of the specification versus conditional logic intended to be part of the specification.

\section[peak]{\peak}
The high-level aim of \peak is to provide a natural object-oriented view of hardware, in which a circuit is defined as a Python class. \peak circuits declare sub-components in their \dunder{init} method\footnote{\dunder{init} is the standard initializer method in Python, which is similar to but not quite equivalent to a constructor in C++.  A more thorough explanation can be found in the Python reference manual~\cite{pyinit}.} and define their behavior in their \dunder{call} method.\footnote{\dunder{call} overloads the function call syntax, i.e., \code{foo(args)} $\equiv$ \code{foo.__call__(args)}} A circuit's inputs are the arguments to its call method and its outputs are the return values of the method. Sub-components are included simply by calling them as functions.  

\begin{examplec}
\begin{lstlisting}
@family_closure(family_group) # 1
def gen(family): # 2
    BV = family.BitVector    
    T = BV[8]                    
    Bit = family.Bit                 
    Register = family.gen_register(T, 0)
    
    @family.compile(locals(), globals()) # 3
    class ALU(Peak): # 4
        def __call__(self, 
                op: Bit, in_0: T, in_1: T) -> T: # 5
            if op: 
                return in_0 + in_1
            else: 
                return in_0 * in_1
            
    @family.compile(locals(), globals()) # 6
    class RegALU(Peak): # 7
        def __init__(self):
            self.alu = ALU()
            self.reg_0 = Register()
            self.reg_1 = Register()
            
        def __call__(self, 
                instr: BV[2], in_0: T, in_1: T) -> T: 
            op = instr[0]
            acc = instr[1]
            out = self.alu(
                op, self.reg_0, self.reg_1
            )
            if acc: 
                self.reg_0 = out
            else: 
                self.reg_0 = in_0
            self.reg_1 = in_1
            return out
    return RegALU
\end{lstlisting}
\vpre
\caption{Peak code.}
\label{code:peak}
\vpost
\end{examplec}

In Example~\ref{code:peak}, we show a small example of \peak code. Code points of interest have been annotated with \code{# n}.  We start by explaining \code{ALU} (\code{# 4}) and \code{RegALU} (\code{# 7}); then, in Section \ref{sec:peak:int} we discuss the remaining code points. The \code{ALU} class performs either an add or a multiply on two data inputs (\code{in_0}, \code{in_1}) and is controlled by a single bit \code{op}. We show the results of compiling this ALU to Verilog using the MLIR~\cite{mlir} backend to \magma in Example~\ref{code:verilog}.

The \code{RegALU} class instantiates an \code{ALU} and two \code{Register}s. \code{RegALU} is controlled by a two-bit signal \code{instr}, where bit 0 is the \code{ALU} op and and bit 1 is an \code{acc} flag.  \code{RegALU} passes the contents of its registers to the \code{ALU} and outputs the \code{ALU}'s output.  When the \code{acc} flag is set, it stores its output in \code{reg_0}; otherwise, it stores the first input. The observant reader will note that the registers in Example \ref{code:peak} are not called as functions.  Instead, they are simply read and written as instance attributes.  We provide this syntax to allow registers' next state to be dependent on current state (which is impossible with the \dunder{call} syntax).

\subsection{\peak Normal Form} The \astt library is used to convert a \peak program to a \hwt program.  This is achieved by first performing a typical single static assignment (SSA) transformation~\cite{ssa1988}, i.e., introducing unique variables for every assignment and replacing control flow with \texttt{phi} statements.  Next, all \code{return} statements are replaced with assignments to fresh identifiers.  Next, the bodies of \code{if} blocks are inlined into their enclosing blocks, and phi nodes are replaced with \code{ite} calls (a method on the primitive type \code{Bit}).  Finally, we construct the return value by reconstructing the condition structure in a nested \code{ite}.  In this form, the program is a pure \hwt program.  The transformed \peak code for \code{ALU.__call__} in Example \ref{code:peak} is shown in Example~\ref{code:normal}. 

Special care is needed to handle attribute writes (e.g., registers) as they do not behave like other names.  At a high level, the compiler simply generates a fresh name for each written attribute which is initialized at the top of the program.  Next, it replaces all references to the attribute with references to the fresh name.  Finally, it writes the generated name back to the attribute at the end of the program.  

The existence of multiple returns complicates this basic scheme, as there are multiple ``ends'' of the program. Hence, at each return location, the state of each attribute (i.e., the value held in the attribute's associated name) is stored in a ``final'' name, so that the proper value may be written to the attribute at end of the program. Then, at the end of the program the final names are multiplexed, in a similar matter to the rebuilding of return values, before being written back. We show the transformation of the simple counter shown in Example~\ref{code:counter} in Example~\ref{code:counter:normal}.

%As an alternative to the function call syntax \peak allows types to define \cdnote{I dont know how to say what I am trying to say concisely.  Consider \ref{code:ideal} note how the registers are read by reading the attribute to which they were initially assigned (in python names are references not values) and how they are updated by assigning to that attribute. This is not a special case for registers a type may describe what it means for it to be read and written instead of describing its self as function of its inputs}

%A \peak program without any transformations or alterations \textit{is} a Python program and hence is executable. However, in order to generate a formal model or RTL, the program must be transformed into a single basic block in SSA form. In other words, \code{if} statements must be inlined, and the function must have a single \code{return} statement.  This is achieved by first performing the SSA pass mentioned above and replacing all \code{return} statements with assignments to fresh identifiers.  Next, the body of \code{if} blocks are inlined into the enclosing blocks, and phi nodes are replaced with \code{ite} (if-then-else) calls (a method on the primitive type \code{Bit}).  Finally, we construct the final return value by reconstructing the condition structure in a nested if-then-else expression. In this form, the program is a pure \hwt program.  For example, \code{ALU.__call__} in \ref{code:ideal} would be transformed to the code shown in Example~\ref{code:normal}.  

\begin{examplec}
\begin{lstlisting}[language=verilog]
module ALU(
    input op,
    input [7:0] in_0, in_1,
    input CLK, ASYNCRESET,
    output [7:0] O
    );
    
    wire [1:0][7:0] _GEN = {
        {in_0 + in_1}, {in_0 * in_1}
    };
    assign O = _GEN[op];
endmodule
\end{lstlisting}
\vpre
\caption{ALU compiled to Verilog using the MLIR backend of \magma.}
\label{code:verilog}
\vpost
\end{examplec}

\begin{examplec}
\begin{lstlisting}
class ALU(Peak):
    def __call__(self, 
            op: Bit, in_0: T, in_1: T) -> T:
        cond_0 = op
        r_val_0 = in_0 + in_1
        r_val_1 = in_0 * in_1
        r_val_f = cond_0.ite(r_val_0, r_val_1)
        return r_val_f
\end{lstlisting}
\vpre
\caption{ALU in \peak normal form as generated by the compiler modulo a slight simplification of generated names.}
\label{code:normal}
\vpost
\end{examplec}

\begin{examplec}
\begin{lstlisting}
@family.compile(locals(), globals())
class Counter(Peak):
	def __init__(self):
		self.reg = Register()

	def __call__(self, en: Bit, rst: Bit) -> T:
		if rst:
			self.reg = T(0)
			return T(0)

		if en:
			state = self.reg
			if state < MAX_COUNT - 1:
				next_state = state + 1
			else:
				next_state = T(0)
			self.reg = next_state
			return state
		else:
			return self.reg
\end{lstlisting}
\vpre
\caption{A counter with a reset and enable.}
\label{code:counter}
\vpost
\end{examplec}
\begin{examplec}
\begin{lstlisting}
def __call__(self, en: Bit, rst: Bit) -> T:
    self_reg_0 = self.reg
    cond_0 = rst
    self_reg_1 = T(0)
    self_reg_f_0 = self_reg_1
	r_val_0 = T(0)
    cond_2 = en
    state_0 = self_reg_0
    cond_1 = state_0 < MAX_COUNT - 1
    next_state_0 = state_0 + 1
    next_state_1 = T(0)
    next_state_2 = cond_1.ite(
        next_state_0, next_state_1
    )
    self_reg_2 = next_state_2
    self_reg_f_1 = self_reg_2
	r_val_1 = state_0
    self_reg_f_2 = self_reg_0
	r_val_2 = self_reg_0
	self_reg_f = cond_0.ite(
		self_reg_f_0,
		cond_2.ite(self_reg_f_1, self_reg_f_2)
	)
	self.reg = self_reg_f
	r_val_f = cond_0.ite(
		r_val_0,
		cond_2.ite(r_val_1, r_val_2)
	)
	return r_val_f
\end{lstlisting}
\vpre
\caption{A counter in \peak normal form as generated by the compiler.  The names have been simplified and additional line breaks have been inserted to increase readability.}
\label{code:counter:normal}
\end{examplec}
\subsection[peak:adt]{Algebraic Data Types}
%\ready

\peak also supports ADTs. 
%which can be can defined using the syntax shown in \ref{code:adt}. 
As ADTs in \peak must be realizable in hardware, we limit them to finite (non-recursive) types. Beyond the usual benefits of abstraction and type safety, ADTs provide a natural abstraction for ISAs: a sum type can be used to specify categories of instructions with different layouts; and product types can used to define the fields of each layout. Example~\ref{code:peak} uses a single bit to control its operation.  However, by doing so we fix the encoding of the ISA.  Instead, designers can define ISAs as ADTs as shown in Example~\ref{code:isa}. 

Using ADTs to represent ISAs has two main benefits.  First, it allows the decode logic to be modified without modifying the functional specification (i.e., the \dunder{call} method).  For instance, to change the encoding of an ADD instruction from \code{op == 1} to \code{op == 0} in the original example (\ref{code:peak}), we would need to update the line \code{if op:} to \code{if ~op:}. In contrast, in Example \ref{code:isa}, we just need to change the definition of \code{AluOp}.  While these two edits are of similar complexity for the toy examples shown here, the ADT-based specification is much more maintainable for more complex examples, as most of the complexity tends to lie in the \dunder{call} method. The second main benefit of using ADTs to describe ISAs is type safety.  In the original example, it would be possible for a designer to accidentally use bit 0 as the \code{acc} flag and bit 1 as the \code{op}. In contrast, comparing a member of \code{AluOp} to a member of \code{RegCtrl} would lead to a type error.  

% \begin{examplec}
% \begin{center}
% \begin{minipage}{0.30\textwidth}
% \begin{lstlisting}
% class P(Product):
%     field_1 = T1
%     ...
%     field_n = TN
% \end{lstlisting}
% \end{minipage}%
% \hfill
% \begin{minipage}{0.35\textwidth}
% \begin{lstlisting}
% class T(TaggedUnion):
%     variant_1 = T1
%     ...
%     variant_n = TN
% \end{lstlisting}
% \end{minipage}%
% \hfill
% \begin{minipage}{0.30\textwidth}
% \begin{lstlisting}
% class E(Enum):
%     name_1 = val_1
%     ...
%     name_n = val_n
% \end{lstlisting}
% \end{minipage}
% \end{center}
% \vpre
% \caption{Definition of syntax for ADTS}
% \cdnote{candidate for cutting}
% \label{code:adt}
% \vpost
% \end{examplec}

\begin{examplec}
\begin{lstlisting}
class AluOp(Enum):
    ADD = 1
    MUL = 0

class RegCtrl(Enum):
    ACC = 1
    BYPASS = 0

class Inst(Product):
    op = AluOp
    ctrl = RegCtrl
\end{lstlisting}

\begin{lstlisting}
...
@family.compile(locals(), globals()) 
class ALU(Peak):
    def __call__(self,
            op: AluOp, in_0: T, in_1: T) -> T:
        if op == AluOp.Add:
            return in_0 + in_1
        else: 
            return in_0 * in_1

@family.compile(locals(), globals())          
class RegALU(Peak):
    def __init__(self):
        self.alu = ALU()
        self.reg_0 = Register()
        self.reg_1 = Register()

    def __call__(self,
            instr: Inst, in_0: T, in_1: T) -> T:
        out = self.alu(
            instr.op, self.reg_0, self.reg_1
        )
        if instr.ctrl == RegCtrl.ACC: 
            self.reg_0 = out
        else: 
            self.reg_0 = in_0
        self.reg_1 = in_1
        return out
\end{lstlisting}
\vpre
\caption{Defining an ISA as an ADT.}
\label{code:isa}
\vpost
\end{examplec}

%\subsubsection{Assemblers}\label{sec:peak:adt:asm}
%\ready

When ADTs are compiled to hardware, they must be encoded as bitvectors. While \peak provides reasonable defaults for the encoding (e.g., \code{Product} types encoded as the concatenation of their fields), a designer may desire a specific bit-level encoding. \peak provides a simple interface to allow this.

% \peak supports this with the following 4 functions:
% \begin{itemize}
%     \item \code{$\sig{assemble}{ADT, BitVector[n]}$} \\
%     Transforms an ADT constant into a \bv.
%     \item \code{$\sig{from_fields}{[BitVector[m]], BitVetor[n]}$}\\
%     Constructs an assembled ADT from its assembled fields.
%     \item \code{$\sig{extract}{BitVector[n], field, BitVector[m]}$}  \\
%     Extracts a field of an ADT from its \bv representation.
%     \item \code{$\sig{match}{BitVector[n], variant, Bit}$} \\
%     Checks if a \bv representation of a sum or enum matches a variant.
% \end{itemize}

% These functions allow the user to build a behavioral sub-type~\cite{liskov94} of the ADT where construction calls the \code{assemble} functions and operations on the ADT are performed with calls to \code{extract} and \code{match}.  \autoref{rule:adt} shows the transformation of the ADT operations.  

% \begin{table}[]
% \begin{align*}
%     \produces{adt_constant}{assemble(adt_constant)}\\
%     \produces{ProductT(field=val, ...)}{from_fields(field=assemble(val), ...)} \\
%     \produces{product.field}{extract(product_bv, "field")}\\
%     \produces{tagged.varient.value}{extract(tagged_bv, "varient")}\\
%     \produces{tagged.varient.match}{match(tagged_bv, "varient")}\\
%     \produces{enum == enum_t.Op}{match(enum_bv, enum_t.Op)}
% \end{align*}
% \caption{The transformation of ADT operations to functions on bitvectors.}
% \label{rule:adt}
% \end{table}

\subsection[peak:int]{\peak Internals and Extensions}
We now explain the remaining code points in Example~\ref{code:peak}.
We highlight a few simple requirements: \peak classes must inherit from the \code{Peak} class (\code{# 4} and \code{# 7}), and the type annotations in the \code{__call__} method (\code{# 5}) are \emph{not} optional, as they are needed to generate ports in a \magma context.

Code point \code{# 2} constructs a closure around the \code{ALU} and \code{RegALU} classes. It takes a single argument, which is a \emph{family} object.  The family mechanism is the means by which the different interpretations (Python, SMT, \magma) for the same \peak code are provided.  Each family object contains one set of implementations for the primitives used by the constructed module (minimally: \bit, \bv, ADTs, registers).  Note how all types are accessed through the \code{family} object.
\code{family.compile} (\code{# 3} and \code{# 6}) invokes the PEak compiler, passing the current symbols to the compiler with \code{locals(), globals()}.  Each family can define its own compilation flow.  For example, the SMT and \magma families rewrite \dunder{call} code into the \peak normal form.

Finally, the \code{family_closure} decorator (\code{# 1}) takes a single parameter, which associates the decorated closure with a specific \emph{family group}, an object (typically a module) with attributes \code{PyFamily}, \code{SMTFamily}, and \code{MagmaFamily}, providing families with the Python, SMT, and \magma interpretations, respectively.  Default implementations for each family can be obtained by using a specific family group that is included with \peak.  The purpose of an explicit family group parameter is to allow extensions beyond this default implementation.
For example, an extended family group could include a floating point type which wraps verilog IP in a Magma context, uses the \hwt floating point type in a Python context, and constructs an uninterpreted function in an SMT context. 

\subsection[peak:usage]{Verification and Test of \peak Circuits}
One of the goals of \peak is to make functional testing simple and to democratize formal verification by making the experience nearly equivalent to writing functional tests. As an example, we check whether the code in \ref{code:isa} always writes its second input to \code{reg_1}, first using random testing then using formal verification. In \ref{code:test:random}, a python instance of the \code{ALU} is instantiated. Next, all possible instructions are exhaustively generated by iterating over all values of \code{AluOp} and \code{RegCtrl}.\footnote{The inner-loop uses the \code{field_dict} attribute of the \code{RegCtrl} type which returns a \code{dict} (mapping type) of names to enum members allowing programmatic generation of such tests.} Then, the registers are set to random initial states, and random inputs are passed to the \code{ALU}. Finally, we assert the postcondition that \code{reg_1} contains the value of \code{i1}.

In \ref{code:test:formal} we show the formal verification of this property which is similar to the random test. First, free SMT variables for the initial states, inputs, and instruction are constructed.  Then, we set the initial state and execute the circuit.  Finally, we use CVC4~\cite{DBLP:conf/cav/BarrettCDHJKRT11} via pySMT to formally verify that \code{reg_1} contains the value of \code{i1} by asserting the negation of the property.  

\begin{examplec}
\begin{lstlisting}
py_alu = gen.Py()
# iterate over all possible instructions
for alu_op in (AluOp.ADD, AluOp.MUL):
    for reg_mode in RegCtrl.field_dict.values():
        # set initial state to random
        py_alu.reg_0 = random_bv(8)
        py_alu.reg_1 = random_bv(8)
        # use random input variables
        i0 = random_bv(8)
        i1 = random_bv(8)
        instr = Inst(alu_op, reg_mode)
        out = py_alu(instr, i0, i1)
        post_condition = py_alu.reg_1 == i1
        assert post_condition
\end{lstlisting}
\vpre
\caption{Random testing of a \peak circuit.}
\label{code:test:random}
\vpost
\end{examplec}

\begin{examplec}
\begin{lstlisting}
initial_reg_0 = SMTBitVector[8]()
initial_reg_1 = SMTBitVector[8]()
i0 = SMTBitVector[8]()
i1 = SMTBitVector[8]()
instr = make_symbolic(Inst)

smt_alu = gen.SMT()
# set the initial state to be symbolic
smt_alu.reg_0 = initial_reg_0
smt_alu.reg_1 = initial_reg_1
# symoblically execute the circuit
out = smt_alu(instr, i0, i1)
post_condition = to_pysmt(smt_alu.reg_1 == i1)

# pysmt code
with Solver("cvc4") as s:
    s.add_assertion(Not(post_condition))
    if s.solve():
        print("Counter example found")
    else:
        print("Verified")
\end{lstlisting}
\vpre
\caption{Verification of a \peak circuit using the CVC4 backend
of pySMT.}
\label{code:test:formal}
\vpost
\end{examplec}

% \subsubsection{Peak Poke Protocol}
% \cdnote{I think I will cut this section}
% \peak allows types to define the interpretation of assignment and reference, as was case for \code{Register}.   See \ref{code:proto}
% \begin{examplec}
% \begin{lstlisting}
% class Reg: 
%     def _peak_(self):
%         self.current_state

%     def _poke_(self, value):
%         self.current_state = value
% \end{lstlisting}
% \caption{}
% \label{code:proto}
% \end{examplec}
\section[eval]{Evaluation}
%\cdnote{2.25 pages}
\peak has been used in the design of three generations of CGRAs: Garnet~\cite{aha2020dac}, Amber~\cite{ahahotchips}, and our most recent architecture, Onyx. Amber and Onyx were fabricated in 16 nm and 12 nm commercial CMOS technologies respectively. \peak's  unique capabilities have also enabled a number of research projects. Here, we present a summary of results from two of these projects. First, we discuss the CGRA design space exploration framework APEX~\cite{apex}, which uses \peak to generate high-performance RTL.  Second, we describe our work on compiler rewrite rule synthesis~\cite{ross2022}, which uses \peak's formal model to synthesize instruction selection rewrite rules efficiently.
\subsection[eval:dse]{APEX}%\ready
% \begin{itemize}
%     \item Briefly describe DSE project (focus on complexity of the parameter space)
%     \item Show small example of meta programming
%     \item Discuss pipelining to prep model checking
%     \item Results to justify high performance RTL
% \end{itemize}

%In the Agile Hardware Center, we aim to create tools that enable fast and easy design space exploration of interesting accelerator architectures. In particular,

%We have investigated the course-grained reconfigurable arrays (CGRA) as a promising accelerator architecture for image processing and machine learning applications. The architecture of a CGRA processing element (PE) directly affects the performance and energy efficiency of the applications running on the accelerator. \peak enables design space exploration of CGRA PEs, and we have automated this design space exploration in APEX~\cite{apex}.

APEX aims to automatically specialize a CGRA's processing element (PE) architecture to an application or a class of applications. First, it uses frequent subgraph mining and analysis techniques to find common computational patterns in applications of interest. After finding frequent subgraphs, APEX merges these graphs into a new graph. This new graph acts as a specification of a specialized PE architecture capable of accelerating the applications.

APEX considers three axes while specializing PEs: number and type of operations within the PE, intraconnect within each PE, and number of inputs and outputs to and from the PE. Each has a direct effect on the complexity and capability of the PE and resulting CGRA. It is important to note that this parameter space is beyond the expressive capabilities of Verilog, which, for example, cannot parameterize the number of ports on a module, whereas in PEak, such parameterization is trivial.

After performing this analysis, APEX automatically converts the graph specification of each PE into a PEak program. At this point, APEX automatically inserts pipeline registers into the design to ensure high performance. The meta-programming utilities in \peak, including loop unrolling and if-statement inlining, make this conversion possible.  

\begin{figure*}
    \centering
    \includegraphics[width=0.6\textwidth]{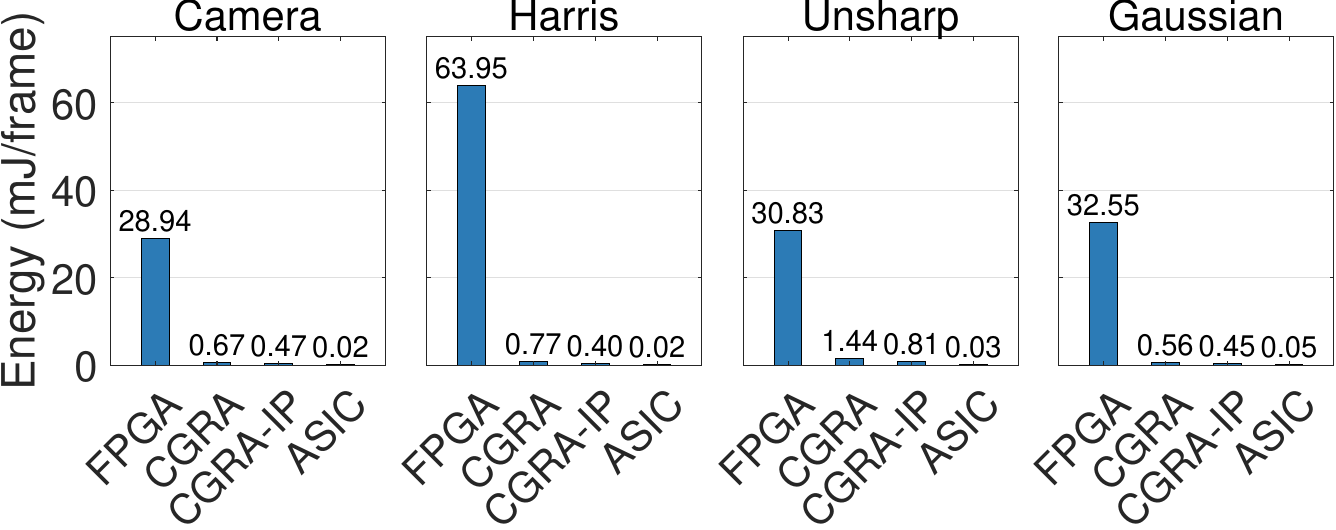} \vspace{-2pt}
    \includegraphics[{trim=0 0 0 0pt},clip,width=0.6\textwidth]{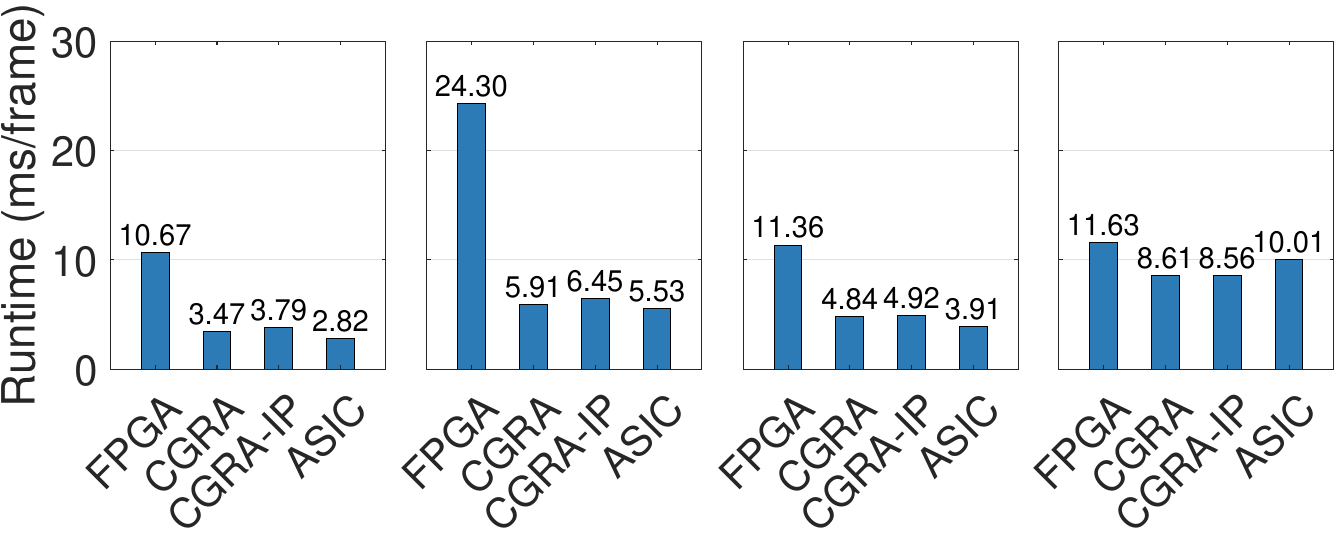}
    \vpre
    \caption{Energy and runtime comparison between an FPGA, an unspecialized CGRA, an APEX-specialized CGRA, and an ASIC. Figure courtesy of Melchert et al.~\cite{apex}.}
    \label{fig:accelerator-comp-ip}
    \vpost
\end{figure*}

Figure~\ref{fig:accelerator-comp-ip} shows the results of evaluating APEX on four image-processing applications: camera pipeline, harris corner detection, unsharp, and gaussian blur.  For each application, we compare an APEX-specialized PE (CGRA-IP) to results obtained using an FPGA, an unspecialized CGRA, and an ASIC. We compare both the energy consumed and the application runtime. The specialized CGRA-IP consumes 18\% to 47\% less energy than a generic CGRA with no specialization, while providing comparable performance.

%The formal model of each PE is used to synthesize an updated compiler, which is necessary to run applications on the new architecture. For each generated PE, the formal model is used to synthesize new rewrite rules for each operation in the target applications. These rewrite rules are used in conjunction with an instruction selection algorithm to form the updated compiler. 

%Depending on the specialization or generalization level of a CGRA PE, the accelerator may have much of the flexibility of an FPGA or much of the energy efficiency and performance as an application specific accelerator (ASIC). To evaluate APEX, we analyze four image processing applications: camera pipeline, harris corner detection, unsharp, and gaussian blur. Next, APEX automatically generates PE-IP which is specialized to all four image processing applications, and CGRA-IP which is a CGRA that contains the specialized PE.

% CGRA-ML achieves 14 $\times$ lower energy than the FPGA implementation, and 36\% lower energy than the baseline CGRA. We also compare CGRA-ML against a state of the art neural network accelerator, Simba \cite{simba}. CGRA-ML approaches the area and energy consumption of Simba, while still being configurable.

% \begin{figure}
%     \centering
%     \includegraphics[width=0.3\textwidth]{figs/ml-energy-resnet.eps} 
%     \includegraphics[width=0.29\textwidth]{figs/ml-runtime-resnet.eps} 
%     \vspace{-3pt}
%     \caption{Comparison of machine learning applications running on an FPGA, the baseline CGRA, CGRA-ML, and Simba.}
%     \label{fig:accelerator-comp-ml}
% \end{figure}

\subsection[eval:synth]{Rewrite Rule Synthesis}%\ready

A working application compiler for each generated PE is required to perform realistic benchmarking of PEs during design space exploration.
During the instruction selection phase of code generation, \emph{rewrite rules} are used to map computations described in an intermediate representation (IR) to concrete inputs, outputs, and instructions on the PE.  Each distinct PE requires its own set of rewrite rules.
Creating these rules manually is both labor-intensive and error-prone.  Furthermore, manual construction would make automatic design space exploration impossible. In a recent work~\cite{ross2022}, we show how these rewrite rules can be efficiently and automatically synthesized, given a formal SMT model of the IR and the target PE. In that work, we conveniently use \peak to describe both, making it easy to extract the SMT models.

%\begin{examplec}
%\begin{lstlisting}
%class IRAdd:
%    def __call__(self, x: BV[16], y: BV[16]) -> BV[16]:
%        return x + y

%\end{lstlisting}
%\vpre
%\caption{Example IR instruction for 16-bit addition} 
%\label{code:iradd}
%\vpost
%\end{examplec}

\newcommand{\IRSub}{\mathsf{bvsub}\xspace}
\newcommand{\IRAdd}{\mathsf{bvadd}\xspace}
\newcommand{\ALU}{\mathit{ALU}\xspace}
\newcommand{\inst}{\mathit{inst}\xspace}

As an example, consider the rewrite rule for a 16-bit subtraction targeting the \code{ALU} described in Example \ref{code:alu}.  The rule specifies that the \code{invert_0}, \code{invert_1}, and \code{op} fields of \code{Inst} should be set to \code{InverterCtrl.ident}, \code{InverterCtrl.invert}, and \code{AluOp.ADD} respectively. Instead of manually creating this rule, it can be synthesized by solving the following SMT query: $\exists\, \inst.\: \forall\, x, y.\: \IRSub(16, x, y) = \ALU(\inst, x, y)$,
where $\IRSub$ is the SMT operator for bitvector subtraction and $\ALU$ is the result of executing the \peak program with the SMT family interpretation
(note that this is a simplified form of the query and does not take into account several complications discussed in \cite{ross2022} such as 
operand ordering, arity mismatches, and state).  We show the construction of this query in ~\ref{code:synth}.

\begin{examplec}
\begin{lstlisting}
class AluOp(Enum):
    ADD = 0
    AND = 1
    OR  = 2

class InverterCtrl(Enum):
    ident  = 0
    invert = 1

class Inst(Product):
    invert_0 = InverterCtrl
    invert_1 = InverterCtrl
    op = AluOp

@family_closure 
def gen(family):
    BV = family.BitVector
    T = BV[8]
    Bit = family.Bit
    @family.compile(locals(), globals()) 
    class ALU(Peak): 
        def __call__(self,
                inst: Inst, in_0: T, in_1: T) -> T:
            if inst.invert_0 == InverterCtrl.invert:
                in_0 = ~in_0
    
            if inst.invert_1 == InverterCtrl.invert:
                in_1 = ~in_1
                cin = Bit(1)
            else:
                cin = Bit(0)
            
            if inst.op == AluOp.ADD:
                res, cout = add_with_carry(
                    in_0, in_1, cin
                )
                return res
            elif inst.op == AluOp.AND:
                return in_0 & in_1
            else:
                return in_0 | in_1
    return ALU
\end{lstlisting}
\vpre
\caption{An ALU supporting 6 operations: Add, Subtract, And, Or, Nand, Nor.}
\label{code:alu}
\vpost
\end{examplec}

\begin{examplec}
\begin{lstlisting}
i0 = SMTBitVector[8]()
i1 = SMTBitVector[8]()
instr = make_symbolic(Inst)

smt_alu = gen.SMT()
# symbolically execute the circuit
out = smt_alu(instr, i0, i1)

# construct the synthesis query (pysmt code)
spec = to_pysmt(out == i0 - i1)
universal_vars = [to_pysmt(i0), to_pysmt(i1)]
query = ForAll(universal_vars, spec)
with Solver("cvc4") as s:
    s.add_assertion(query)
    if s.solve():
        val = s.get_py_value(to_pysmt(instr))
        print("Rule found using instruction:")
        print(disassemble(val))
    else:
        print("No Rule")
\end{lstlisting}
\vpre
\caption{Rewrite rule synthesis query using \peak and pySMT.}
\label{code:synth}
\vpost
\end{examplec}

Another challenge is handling instructions that use compile-time constants such as immediate fields (e.g., add immediate). Using the above formula, we would need a distinct query for each possible compile-time constant. Instead, we can modify the query by finding an instruction that works for every value of the constant, i.e., $\exists\, \inst.\: \forall\, x, y, c.\: \IRAdd(16, x, c) = \ALU(\inst(c), x, y)$.
To solve this query, we want to treat some fields of the instruction as universally quantified and others as existentially quantified. \peak's ability to represent instructions as ADTs makes this possible.

\begin{figure*}
	\centering
	\includegraphics[scale=0.75]{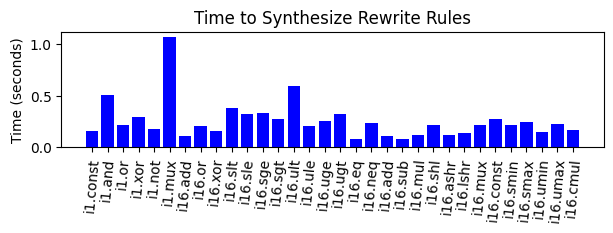}
    \vpre
	\caption{Rewrite rule synthesis times for various IR instructions.}
	\label{fig:cgra-op}
    \vpost
\end{figure*}

Figure~\ref{fig:cgra-op} shows the results of synthesizing rewrite rules for a set of IR instructions.  The maximum time is 1.1 seconds. Synthesizing all of the rules takes less than 30 seconds, fast enough to be used in the loop during design space exploration, and a significant improvement over manual implementation of rules.

\section[related]{Related Work}

The design of \peak draws inspiration from the classic work of Bell and Newell~\cite{isp1970}, which similarly separated the logical description of an ISA from its semantics and bit-level representations.  However, this idea seems to have been largely lost over time and, to our knowledge, is not used in any modern system.  \peak generalizes this idea from ISAs to ADTs.

There are many HDLs designed for general-purpose hardware construction, the most popular being Verilog.  However, Verilog has extremely limited meta-programming capabilities, weak type systems, and poorly defined semantics.  More modern languages with strong type systems like \magma~\cite{truong2019golden} and Chisel~\cite{bachrach2012chisel} ease meta-programming by being embedded in Python and Scala, respectively.  These languages define hardware as a graph of modules which is explicitly wired together.  In contrast, \peak uses an implicit wiring model to avoid combinational loops.  This is a deliberate design decision to keep designs readable and to ensure deterministic behavior.\footnote{This means that certain design patterns that use combinational loops, flip-flops constructed from NAND gates for example, are not expressible in \peak.}  \peak also provides access to a formal model, a feature not available in other HDLs.

\peak is also inspired by Lava~\cite{bjesse1998lava}, a Haskell-based DSL which supports multiple interpretations similar to \peak. Lava programs, like Magma and Chisel programs, describe hardware structurally. C$\lambda$aSH~\cite{baaij2010c} is another Haskell-based DSL which is less structural than Lava. It allows the use of case statements and pattern matching, enabling the construction of complex control structures which are difficult to build structurally. However, it does not have direct support for formal analysis like \peak and Lava. Both of these languages have limited type systems.  In particular, they lack the ADT capability supported by \peak.  Finally, while Haskell is appealing to DSL designers, as it enables elegant meta-programming through the use of type class polymorphism and higher order functions, practice has shown that getting working engineers to adopt a Haskell-based DSL is challenging.

For example, Bluespec SystemVerilog (BSV)~\cite{nikhil2004bluespec}, a term rewriting system (TRS) that describes circuits as a set of guarded atomic actions (rules), originally had a Haskell-like syntax.  However, to appeal to a wider audience, it has since adopted an imperative syntax that is closer to behavioral Verilog. BSV rules describe a circuit's behavior as state updates and outputs predicated on current states and inputs. Abstractly, these rules are atomic and are applied sequentially, one rule at a time.  However, in practice this would lead to extremely inefficient hardware.  Therefore, the BSV compiler attempts to schedule these rules concurrently when possible.  When multiple rules can update the same state element they must be scheduled sequentially.  The choice of schedule can have significant impact on the quality of the resulting hardware. K\^{o}ika~\cite{bourgeat2020essence} is a BSV derivative which aims to eliminate this by giving engineers direct control over the schedule.

A related line of work is high-level synthesis~\cite{daoud2014survey} (HLS) which allows designers to describe the behavior of circuits using a high-level programming language such as C, C++, SystemC, or Matlab. HLS programs describe the algorithmic behavior of a circuit, eschewing low-level details like pipelining and resource allocation. An HLS compiler then determines some minimal set of resources which are capable of performing the described algorithm and an associated schedule of computation, i.e. where and when each operation in the source program takes place.  While HLS is a popular design paradigm and can provide significant engineering efficiency gains, it often produces low-performance RTL~\cite{agarwal2010comparative}.

Contemporary work on ISA specification falls into two main categories: ad hoc specification of existing ISAs~\cite{x86spec2014,armspec2016} and frameworks which are more analogous to \peak for specifying ISAs such as
SAIL~\cite{sail2015}, ILA~\cite{ila2018}, and ISA-Formal~\cite{isaformal2016}.  These systems use declarative descriptions of the semantics of instructions as state updates predicated on the bit-level representation of an instruction.  These are powerful tools, but they cannot be used to generate RTL.  While this disconnect makes sense when verifying new RTL against an existing ISA specification, it is tedious when the ISA itself being developed, as for each new candidate ISA, both its RTL and its specification must be written separately.  In contrast, \peak uses a procedural model in which bit-level encodings are decoupled from the behavioral specification. Further, \peak can be used both for specification and RTL-generation.
\section[conclusion]{Conclusion}
\peak provides designers with the means to specify a single source of truth for hardware design. This has been proven to be a useful paradigm for enabling novel automated design methodologies which incorporate formal methods. These methodologies have enabled us to develop three generations of CGRA architectures.  

\peak is built on top of \hwt and \astt.  \hwt provides a Pythonic interface to functional simulation, formal SMT models, and RTL generation via \magma. \astt provides infrastructure for Python AST analysis and transformations and enables the reinterpretation of Python control flow.  Such tools may be useful for other new languages which, like \peak, aim to provide a single source of truth.

\LetLtxMacro{\section}{\oldsection}
\LetLtxMacro{\subsection}{\oldsubsection}
%\LetLtxMacro{\subsubsection}{\oldsubsubsection}
\bibliographystyle{plain}
\bibliography{ref}
\end{document}